\documentclass[]{spie}  %>>> use for US letter paper
\usepackage{amsmath,amsfonts,amssymb}
\usepackage{graphicx}
\usepackage[colorlinks=true, allcolors=blue]{hyperref}

\title{Performance of Kitt Peak's Mayall 4-meter Telescope During DESI Commissioning}

\author[a]{Aaron M. Meisner}
\author[a]{Behzad Abareshi}
\author[a]{Arjun Dey}
\author[b]{Connie Rockosi}
\author[a]{Richard Joyce}
\author[a]{David Sprayberry}
\author[c]{Robert Besuner}
\author[d]{Klaus Honscheid}
\author[e]{David Kirkby}
\author[d]{Hui Kong}
\author[c]{Martin Landriau}
\author[c]{Michael Levi}
\author[f]{Ting Li}
\author[a]{Bob Marshall}
\author[d]{Paul Martini}
\author[d]{Ashley Ross}
\author[g]{David Brooks}
\author[g]{Peter Doel}
\author[h]{Yutong Duan}
\author[i]{Enrique Gazta\~{n}aga}
\author[j]{Christophe Magneville}
\author[k]{Francisco Prada}
\author[l]{Michael Schubnell}
\author[l]{Gregory Tarl\'{e}}
\affil[a]{NSF's National Optical-Infrared Astronomy Research Laboratory, 950 N. Cherry Ave., Tucson, AZ 85719, USA}
\affil[b]{Department of Astronomy and Astrophysics, University of California Santa Cruz, 1156 High St., Santa Cruz, CA 95064, USA}
\affil[c]{Lawrence Berkeley National Laboratory, Berkeley, CA 94720, USA}
\affil[d]{Department of Physics, The Ohio State University, Columbus, OH 43210, USA}
\affil[e]{Department of Physics and Astronomy, University of California, Irvine, CA 92697, USA}
\affil[f]{Observatories of the Carnegie Institution for Science, 813 Santa Barbara St., Pasadena, CA 91101, USA}
\affil[g]{Department of Physics \& Astronomy, University College London, Gower Street, London, WC1E 6BT, UK}
\affil[h]{Physics Department, Boston University, Boston, MA 02215, USA}
\affil[i]{Institute of Space Sciences (ICE, CSIC), Campus UAB, Carrer de Can Magrans, s/n, 08193 Bellaterra (Barcelona), Spain}
\affil[j]{DSM/Irfu/SPP, CEA-Saclay, F-91191 Gif-sur-Yvette Cedex, France}
\affil[k]{Instituto de Astrof\'{i}sica de Andaluc\'{i}a (CSIC), Glorieta de la Astronom\'{i}a, E-18080 Granada, Spain}
\affil[l]{University of Michigan, Randal Laboratory, 450 Church Street, Ann Arbor, MI 48109, USA}

\begin{document} 
\maketitle

\begin{abstract}
In preparation for the Dark Energy Spectroscopic Instrument (DESI), a new top end was installed on the Mayall 4-meter telescope at Kitt Peak National Observatory. The refurbished telescope and the DESI instrument were successfully commissioned on sky between 2019 October and 2020 March. Here we describe the pointing, tracking and imaging performance of the Mayall telescope equipped with its new DESI prime focus corrector, as measured by six guider cameras sampling the outer edge of DESI's focal plane. Analyzing $\sim$500,000 guider images acquired during commissioning, we find a median delivered image FWHM of 1.1 arcseconds (in the r-band at 650 nm), with the distribution extending to a best-case value of $\sim$0.6 arcseconds. The point spread function is well characterized by a Moffat profile with a power-law index of $\beta \approx 3.5$ and little dependence of $\beta$ on FWHM. The shape and size of the PSF delivered by the new corrector at a field angle of 1.57 degrees are very similar to those measured with the old Mayall corrector on axis. We also find that the Mayall achieves excellent pointing accuracy (several arcseconds RMS) and minimal open-loop tracking drift ($< 1$ milliarcsecond per second), improvements on the telecope’s pre-DESI performance.  In the future, employing DESI’s active focus adjustment capabilities will likely further improve the Mayall/DESI delivered image quality. 
\end{abstract}

\keywords{Mayall Telescope, Kitt Peak National Observatory, DESI}

\section{INTRODUCTION}
\label{sec:intro}  % \label{} allows reference to this section

The Dark Energy Spectroscopic Instrument\cite{Levi2013} (DESI) is a Stage IV ground-based dark energy experiment that will produce an unprecedented three-dimensional map of the Universe. DESI employs the baryon acoustic oscillation (BAO) technique to provide state-of-the-art cosmological constraints to redshifts $>$ 2 by observing several tracer galaxy populations\cite{DESI_Part1}. In total, approximately 35 million redshifts will be obtained by DESI during its five-year survey which will cover nearly the entire northern ($\delta > -30^{\circ}$) sky at high Galactic latitudes\cite{Dey} ($\sim$15,000 square degrees in total). Importantly, DESI will collect redshifts for $\sim$15 million faint ($g \sim 23$ AB) emission line galaxies at $z \sim 1-1.5$, a redshift range largely unexplored with BAO.

% cite Dey Legacy Surveys overview paper in above paragraph?

DESI consists of a next-generation multi-object spectroscopy instrument\cite{DESI_Part2, Martini_DESI_SPIE} installed at Kitt Peak National Observatory's Nicholas U. Mayall 4-meter Telescope in southern Arizona. DESI's ten spectrographs combine to acquire 5,000 spectra simultaneously, spanning the 360-980 nm wavelength range with resolution $\lambda$/$\Delta\lambda$ $\sim$ $2,000-5,500$. DESI's focal plane resides at the prime focus of the Mayall telescope, where a new DESI top end\cite{Gaston_top_end}  has been installed. The DESI corrector\cite{Miller_corrector} provides a large 3.2$^{\circ}$ diameter field of view, of which $\sim$7.5 square degrees is instrumented for spectroscopy. A hexapod allows for fine-grained adjustments of the corrector barrel position. 5,000 fiber positioning robots\cite{Silber_robots,Leitner_positioners} patrol the focal plane to align stars and galaxies with fiber-optic cables connected to the spectrographs.

DESI installation\cite{Allen_installation} completed in 2019 October, and the Mayall facility will be entirely dedicated to DESI operations until the survey's completion. It is therefore important to validate that the Mayall telescope with DESI top end meets DESI's technical requirements while performing at the same high level as did the pre-DESI configuration. In this work, we use DESI commissioning observations to validate the  Mayall telescope's performance in terms of pointing, tracking and delivered image quality. In $\S$\ref{sec:cmx_review} we provide an overview of DESI commissioning. In $\S$\ref{sec:pointing_tracking} we characterize the Mayall/DESI pointing and tracking. In $\S$\ref{sec:diq} we study the Mayall/DESI delivered image quality. We conclude in $\S$\ref{sec:conclusion}.
% could elaborate on the "We conclude" final sentence by mentioning that in the conclusion we comment on further data analysis and integration/testing that could further augment/improve the results presented in this current work.

% faintness of targets - G ~ 23-24 - need good image quality to get enough signal, and also for guiding during long exposures
% wide-field corrector : challenge to get excellent image quality over entire FOV

%phases : installation -> commissioning -> survey validation -> survey operations

% list out the paper's sections as usual in final sentence of intro

\section{DESI Commissioning}
\label{sec:cmx_review}

% find a way to cite Behzad's TCS SPIE paper

% mention protodesi at all????

The DESI instrument is comprised of many component subsystems, making the commissioning process a multifaceted endeavor. DESI commissioning consisted of two distinct phases. First, the DESI Commissioning Instrument\cite{Ross_CI} (CI) campaign provided initial checks on the Mayall/DESI setup, especially the new DESI corrector, during on-sky nighttime observations from 2019 March through 2019 May. The CI hardware was a temporary focal plane instrumented with five commercial CCD imaging sensors meant to mimic DESI's guider cameras. The CI focal plane contained no fiber positioners and hence the run did not include any spectroscopic tests. Nevertheless, the CI campaign resulted in many advancements, including a Mayall/DESI pointing model, a look-up-table for best focus/alignment as a function of telescope orientation, testing of the DESI instrument control software system\cite{Honscheid_ICS} (ICS) and observer interface, validation of the DESI corrector's ability to provide sharp images across the field of view, and a Mayall/DESI field rotation model.

\begin{figure} [ht]
   \begin{center}
   \begin{tabular}{c} %% tabular useful for creating an array of images 
   \includegraphics[height=10cm]{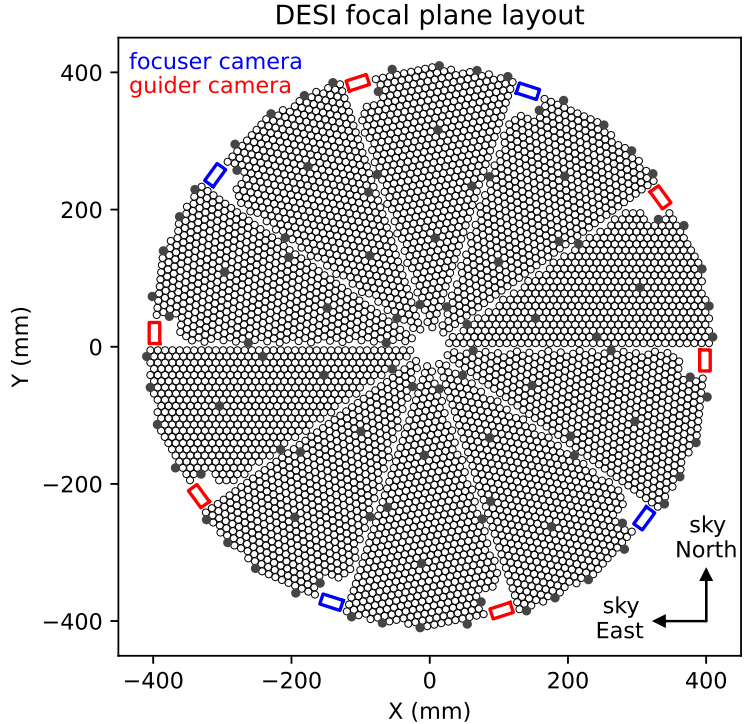}
   \end{tabular}
   \end{center}
   \caption[example] 
%>>>> use \label inside caption to get Fig. number with \ref{}
   { \label{fig:fp_layout} Schematic diagram illustrating the layout of DESI's focal plane. Open black circles are robotic fiber positioners, gray filled circles are fixed reference ``fiducials'', blue rectangles are outlines of the four focuser cameras, and red rectangles are outlines of the six guider cameras. The $\sim$5,000 DESI fiber positioners instrument a solid angle of $\sim$7.5 square degrees. The DESI focal plane angular diameter is 3.2$^{\circ}$, with the guider and focuser cameras centered at 1.57$^{\circ}$ off-axis. Each guider sensor covers a sky area of 24.6 square arcminutes. Source centroid and morphology measurements from DESI's six guider cameras enable the telescope performance assessments in this study.}
   \end{figure}

Following an observing hiatus during the summer of 2019, DESI focal plane commissioning began with spectroscopic first light on 2019 October 22. Unlike the CI run, focal plane commissioning observations were performed with the production DESI focal plane installed at the Mayall and connected to the DESI spectrographs. A schematic diagram of the DESI focal plane layout is shown in Figure \ref{fig:fp_layout}. The DESI focal plane consists of ten wedge-shaped ``petals''. Each petal contains $\sim$500 fiber positioner robots, $\sim$10 ``fiducials'' (fixed references used for metrology), and a `Guide, Focus and Alignment' (GFA\cite{GFA_SPIE}) imaging sensor near the outer edge. DESI employs two distinct types of GFA cameras: guiders and focusers. Guiders are designed such that their images are in focus when incoming light is focused at the positioner fiber tips. On the other hand, focusers are intentionally out of focus relative to the guiders and fiber tips by $\pm$1,500$~\mu$m (one half of each focuser chip is intra focal and the other half is extra focal). During DESI survey observing, the six guider cameras are used for field acquisition and guiding throughout DESI's relatively long ($\sim$20 minute) spectroscopic exposures. The focuser cameras are used for wavefront sensing that allows for DESI to actively monitor and control focus during exposures\cite{Roodman_donuts}.
% in this last sentence cite +/- 30 microns target accuracy of active focus adjustments?

% unique CI feature of imaging the **center** of the focal plane (imaging on-axis)

% Note: If compiling with LaTeX+dvipdf, please ensure images generated from 
% other software packages have their bounding boxes set correctly.

\subsection{DESI Guider Cameras}
\label{sec:gfa}

\begin{figure} [ht]
   \begin{center}
   \begin{tabular}{c} 
   \includegraphics[height=6cm]{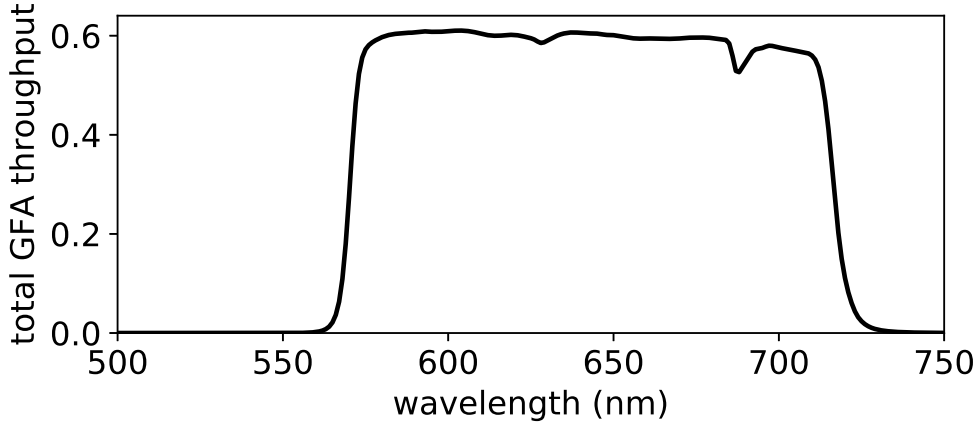}
	\end{tabular}
	\end{center}
   \caption[example] 
   { \label{fig:gfa_total_throughput} GFA throughput curve, taking into account the contributions of atmospheric transmission (at zenith), the Mayall primary mirror reflectance, the DESI corrector throughput, vignetting, the GFA r-band filter and the e2v CCD230-42 quantum efficiency.}
   \end{figure}

Because the Mayall telescope performance assessment throughout the remainder of this study relies on source centroid/morphology measurements from the DESI guider cameras, we provide additional GFA specification details in this subsection. Each DESI GFA camera is an e2v CCD230-42 back illuminated scientific CCD sensor. Each such sensor is 2048 $\times$ 2064 pixels in extent. However, only 2048 $\times$ 1032 pixels of each GFA camera are used by DESI to collect light; the other half of each sensor is masked to enable a frame transfer mode with expedited readout. Each GFA pixel is 15~$\mu$m on a side, corresponding to an angular size of $\Omega_{\rm pix}^{1/2}$ = 0.205$''$ at the outer edge of the DESI focal plane. The angular dimensions of each GFA camera on sky are 7.3$'$ $\times$ 3.4$'$, covering a solid angle of 24.6 square arcminutes. Each guider camera is equipped with a 5 mm thick SDSS r-band filter. Figure \ref{fig:gfa_total_throughput} shows the GFA total throughput curve as dictated by the atmospheric transmission (at zenith), primary mirror reflectance, corrector throughput, vignetting, r-band filter transmission and sensor quantum efficiency. The central wavelength is $\lambda \approx 645$ nm, with a half-width of roughly 75 nm.

The DESI GFA cameras are operated at a typical sensor temperature of $\sim$11$^{\circ}$ Celsius. At this temperature, their dark current is rather large ($\sim$40 e$-$/second/pixel) with strong spatial structure across each sensor. In order to measure source properties as accurately as possible, it is therefore important to detrend the raw guider camera data. Thus, prior to making any of the centroid or morphology measurements used throughout this work, we apply an off-line DESI GFA reduction pipeline to detrend the raw guider images. This detrending includes bias subtraction, dark current subtraction, and flatfielding.

% gain ~3.7 e-/ADU

% mention the gain? readnoise

\subsection{Observing Campaign}
\label{sec:observing}

The DESI focal plane commissioning observing campaign took place from 2019 October 22 through 2020 March 15. Our commissioning campaign achieved its goal of demonstrating that the full DESI system is capable of executing standard DESI survey observations, which consist of long ($\sim$20 minute) guided spectroscopic exposures yielding thousands of faint galaxy redshifts per pointing. However, many of the DESI commissioning observations gathered deviate in a variety of ways relative to the standard DESI exposure sequence. Given the numerous special commissioning tests performed, the resulting data set is very heterogeneous.

\begin{figure} [ht]
   \begin{center}
   \begin{tabular}{c} 
   \includegraphics[height=11cm]{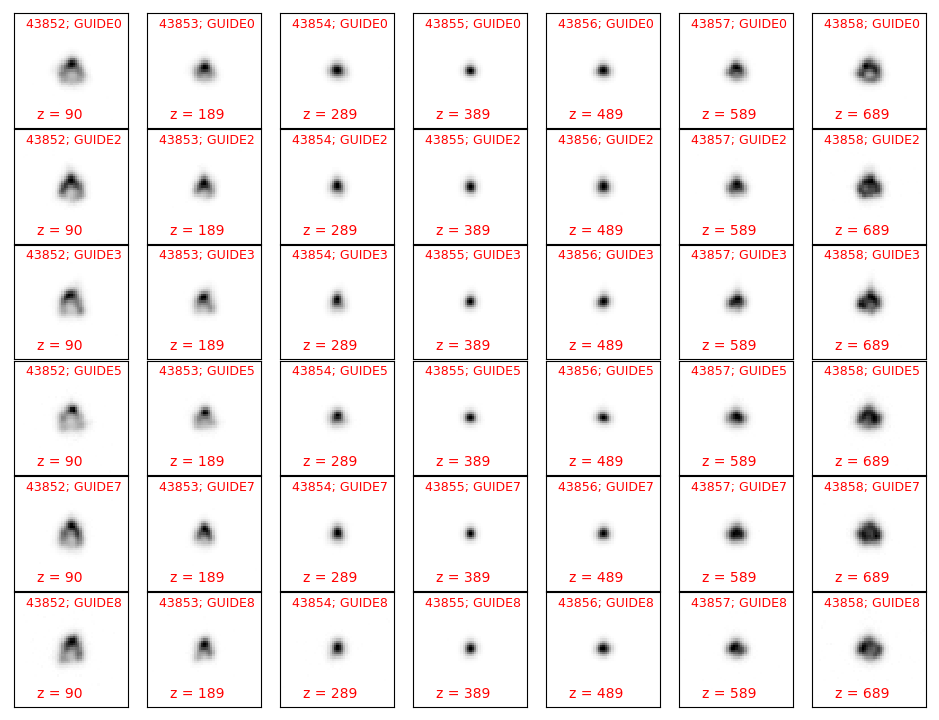} \\
   \includegraphics[height=6cm]{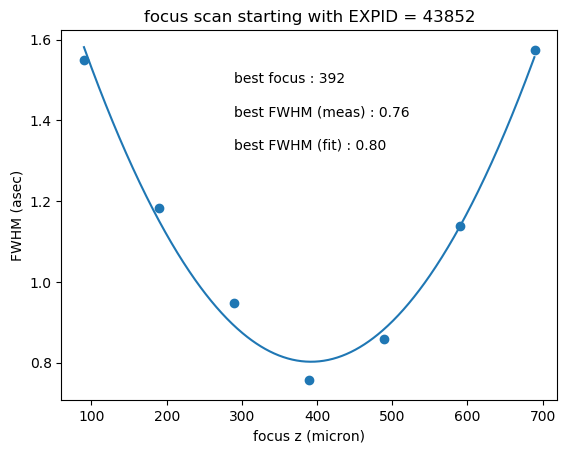}
	\end{tabular}
	\end{center}
   \caption[example] 
   { \label{fig:focus_scan_example} Example of a focus scan used during DESI commissioning to place the guider cameras in focus. Top: One PSF postage stamp per guider camera (row) per exposure (column). Each postage stamp is $10'' \times 10''$. Guider camera names, hexapod z values, and exposure numbers are provided as red annotations. Bottom: Fit of the per-exposure FWHM trend as a function of hexapod z, which determines the z value that best focuses the guider cameras for subsequent observations.}
   \end{figure}

One significant aspect of DESI commissioning observations that differs from production survey operations is that we did not enable active focus adjustments while guiding/exposing. Instead, we determined the best focus by performing focus scans like that illustrated in Figure \ref{fig:focus_scan_example}, where we stepped the distance (z) between the DESI corrector/focal plane and primary mirror in increments of 100~$\mu$m. We then fit a parabola to the resulting trend of guider camera FWHM versus z in order to determine the optimal focus (z value of the parabola minimum). Within the sequence of guider camera focus scan postage stamps shown in Figure \ref{fig:focus_scan_example}, each cutout has an angular extent of $\sim$10$''$ on a side.

The vast majority of the existing DESI guider camera images were acquired while guiding, during which one 5 second GFA exposure is obtained every 10 seconds (the GFA readout time in this operating mode is 5 seconds). These 5 second guider camera frames were typically acquired either during dedicated guiding tests or guided spectroscopic exposures. For both of these applications it is important that the telescope be well-focused, making this large data set of 5 second guiding exposures very valuable for studies of delivered image quality ($\S$\ref{sec:diq}).

\section{Pointing and Tracking}
\label{sec:pointing_tracking}

In preparation for DESI installation, the Mayall telescope control system (TCS) recently underwent a significant upgrade\cite{Sprayberry_Mayall_TCS,Abareshi_Mayall_TCS}. This effort yielded excellent pointing and tracking performance for the Mayall telescope with its pre-DESI top end. Prior to TCS modernization, the Mayall's  pointing accuracy was $\sim$10-20$''$ RMS. Following the TCS enhancements, the pointing accuracy improved to just $\approx$3$''$ RMS. The tracking drift also improved from $\sim$0.5$''$/minute to $\sim$0.04$''$/minute.
% re: previous sentence - ideally would be good to get clarification from Dick as to whether the RMS values quoted in DESI-1421 are per coordinate or 2D

It is important to re-measure these telescope performance metrics following DESI top end installation for several reasons, including: (1) the DESI top end has a different mass and moment than the foregoing configuration; (2) verifying that no other aspects of the new DESI hardware (e.g., the fiber cabling) have hindered the Mayall's pointing/tracking capabilities; (3) checking that no incidental regressions (e.g., loss of encoder cleanliness) have taken place during the extensive DESI installation process.

Minimizing the pointing model's RMS accuracy is, by itself, not especially critical for DESI survey operations. Upon slewing to a new field, DESI's PlateMaker\cite{DESI_Part2} software is capable of detecting pointing corrections of several arcminutes during its astrometric registration process. Still, constructing a high-quality pointing model is a key step toward ensuring excellent tracking performance, which is very important to DESI's success. DESI's long spectroscopic exposures necessitate excellent guiding ($\sim$100 mas RMS), which in turn requires good tracking performance so as to avoid the introduction of excessive astrometric jitter on short timescales.

The DESI CI was constructed to have the same mass and moment as the full DESI focal plane system\cite{Ross_CI}, allowing us to build a Mayall/DESI pointing model during the CI observing campaign. We then applied this same pointing model during DESI focal plane commissioning. On 2020 January 16 we conducted a pointing model observing sequence using the CI pointing model. A pointing model observing sequence consists of a series of long slews all around the observable sky to acquire stars with well known astrometric positions. Typically, the goal is to visit $\sim$30 sky locations spanning a wide range of hour angles and declinations, with a long slew separating each pair of consecutive sky locations. Upon completion of each long slew, a pointing correction is calculated, applied and logged. The resulting set of $\sim$30 pointing offsets can then be used to further refine the pointing model parameters if necessary. The DESI hexapod and atmospheric dispersion compensator (ADC) are kept fixed during pointing model observing sequences, since motions of these components between sky locations can introduce relative shifts in the field centers obtained.

\begin{figure}
   \begin{center}
   \begin{tabular}{c} %% tabular useful for creating an array of images 
   \includegraphics[height=6.5cm]{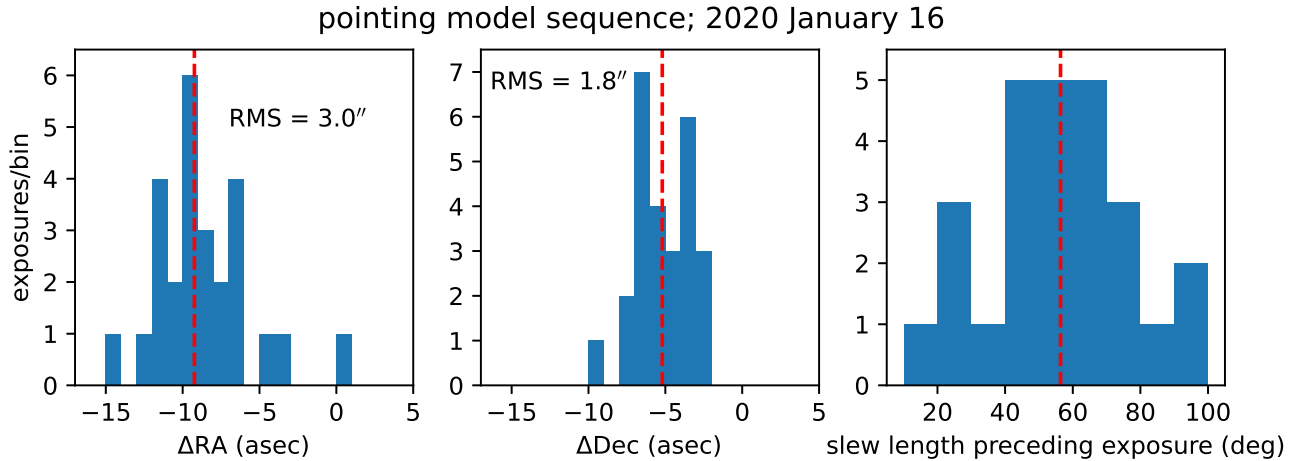}
   \end{tabular}
   \end{center}
   \caption[example] 
%>>>> use \label inside caption to get Fig. number with \ref{}
   { \label{fig:pointing_hist} Summary of a pointing model sequence from DESI observing night 2020 January 16. Left: histogram of RA direction residuals between the commanded telescope position following a long slew and the actual telescope position. Center: histogram of Dec direction residuals between commanded and actual telescope positions during this pointing model sequence. Right: histogram of slew lengths preceding the exposures that contribute to the left and center histograms. The median slew length is quite long, 56$^{^{\circ}}$. Dashed red vertical lines indicate the median value in each panel.}
   \end{figure}

Figure \ref{fig:pointing_hist} shows the results of our 2020 January 16 pointing model observing sequence. The median slew length between consecutive sky locations was 56$^{\circ}$, and the declination ranged from $-35^{\circ}$ to $+75^{\circ}$ while the hour angle spanned from $-71^{\circ}$ to $+73^{\circ}$. The RMS pointing offset following long slews was only 3.0$''$ in the RA direction and 1.8$''$ in the Dec direction. These RMS accuracy values are consistent with those of the pre-DESI Mayall pointing model, demonstrating that the Mayall/DESI configuration can perform to this same high standard. Note that the small systematic shifts of the RA and Dec pointing offset components ($-9''$ and $-5''$, respectively) are merely definitional. In this work we use the \verb|desimeter| software\footnote{\url{https://github.com/desihub/desimeter}} to calculate field centers whereas the pointing model was created using the PlateMaker software. In late January 2020, the PlateMaker software was updated to use a definition of the DESI field center consistent with that adopted by \verb|desimeter|.

Tracking tests are performed by simply remaining at a single sky location for $\sim$10$-$30 minutes while acquiring a series of DESI guider camera images to monitor the motions of stellar source centroids. Again, it is important to keep both the hexapod and ADC lenses fixed during such a sequence, since motions of these components can introduce field center shifts not related to movement of the telescope itself. On 2020 January 28 we performed a 12 minute tracking test which consisted of a series of 20 second DESI guider camera exposures. Analyzing the guider camera source centroids, we find a drift rate of 20 milliarcseconds/minute (13 milliarcseconds/minute) along the RA (Dec) direction. These rates are similar to the previously quoted benchmark values for pre-DESI Mayall tracking drift, validating the Mayall/DESI tracking drift performance.

The Mayall's slew and settle times with its new DESI top end are identical to the pre-DESI Mayall slew and settle times. The Mayall/DESI configuration remains capable of pointing `beyond the pole' (and has done so on a few occasions during DESI commissioning), but DESI does not plan to perform spectroscopic survey observations beyond-the-pole.

% would be good to double check the exposure time value of the 20200128 tracking test exposures used -- not 100% sure the currently stated value of 20 seconds is correct

\section{Delivered Image Quality}
\label{sec:diq}

Delivered image quality (DIQ) is a critical metric for assessing the performance of a telescope plus instrument combination. For DESI, attaining good DIQ is necessary to confine a sufficient fraction of each target's flux within the small 107~$\mu$m ($\approx$1.5$''$) diameter aperture of its associated fiber. The DIQ is determined by a variety of effects, including the atmospheric seeing, turbulence within the telescope dome (`dome seeing'), temperature differential between the primary mirror and ambient, corrector blur and so forth. The Kitt Peak site features excellent atmospheric seeing, and the Mayall telescope was upgraded roughly two decades ago with the specific purpose of further optimizing DIQ\footnote{\url{https://www.noao.edu/noao/noaonews/sep99/node39.html}}. Additionally, the Mayall dome's exterior was re-painted with LO-MIT in 2016, to decrease the nighttime dome versus ambient temperature differential and associated turbulence. During DESI commissioning, observers no longer used the Mayall's `upper' control room, which has a wall that abuts the dome's interior; the corresponding small reduction in heat deposited into the dome may have had a slight positive effect on the DIQ. The DIQ of the Mayall/MOSAIC system that preceded DESI has been thoroughly characterized\cite{dey_valdes}. Mayall/MOSAIC provided a median R-band DIQ of FWHM = 1.17$''$, and the stellar profiles were best described by a Moffat function\cite{Moffat1969} of power law index $\beta = 3.5$, with $\beta$ essentially independent of FWHM\cite{dey_valdes}. Here we seek to compare the DESI r-band DIQ against these Mayall/MOSAIC FWHM and $\beta$ benchmarks.
%since these benchmarks were adopted as requirements during the DESI survey planning process.

The first step toward characterizing the Mayall/DESI DIQ is selecting a large sample of in-focus guider camera images spanning many nights throughout the DESI commissioning campaign. As mentioned in $\S$\ref{sec:observing}, images acquired while guiding provide a natural data set for DIQ studies, since it is reasonable to assume that the telescope would have been well-focused by observers prior to initiating guiding sequences. This full data set consists of 552,076 single-camera guider images. We begin by attempting to construct a pixelized point spread function (PSF) model for each of these full-frame, single-camera guider images. These PSF models are built by combining postage stamps of individual sources and do not assume any functional form or include any analytic component. We use our own custom Python code to create PSF postage stamps, rather than adopting any off-the-shelf PSF-making software. In brief, our PSF-making code gathers $10.5' \times 10.5'$ cutouts centered on well-detected (signal-to-noise $\ge$ 20) stars within each detrended single-camera guider image, subtracts off a sky background level determined within a 5.2$'$-7.4$'$ (radius) annulus surrounding each source, normalizes each source's background-subtracted cutout based on its 1.5$''$ radius aperture flux, applies subpixel shifts to precisely align the centroids of all sources, and then median filters the resulting image cube of normalized/recentered cutouts to arrive at a final PSF postage stamp. Ideally, we would use a weighted mean rather than a median filter (we will likely attempt to implement such an upgrade in the future), but the DESI guider cameras suffer from analog-to-digital conversion issues that create severe hot/cold pixels appearing in unpredictable locations from one image to the next, which makes averaging perilous.

In some cases no astronomical sources are detected and thus no PSF model is generated. In other cases, only a very small number of sources contribute to an image's PSF model. For our FWHM and $\beta$ analysis, we only retain guider images with PSF models built from at least three high-significance sources. We additionally reject highly asymmetric PSF models. Asymmetry is taken to be a proxy for one of several undesirable situations, such as out-of-focus guiding data or streaked images where telescope tracking was lost while exposing. Finally, we remove guider images whose prescan/overscan regions show indications of problematic analog-to-digital conversion, a behavior that sometimes corrupts the DESI GFA readout. After all of these cuts, the resulting sample contains 479,755 single-camera guider images. In selecting this sample, we attempted to avoid selection cuts that would significantly bias the distribution of FWHM values. The median (mean) number of sources contributing to each single-camera PSF cutout is 10 (14.6).

% add footnote to IDLUTILS ATV for radial profile code reference
% http://www.sdss3.org/svn/repo/idlutils/tags/v5_5_5/pro/plot/atv.pro

\begin{figure} [ht]
   \begin{center}
   \begin{tabular}{c} 
   \includegraphics[height=8cm]{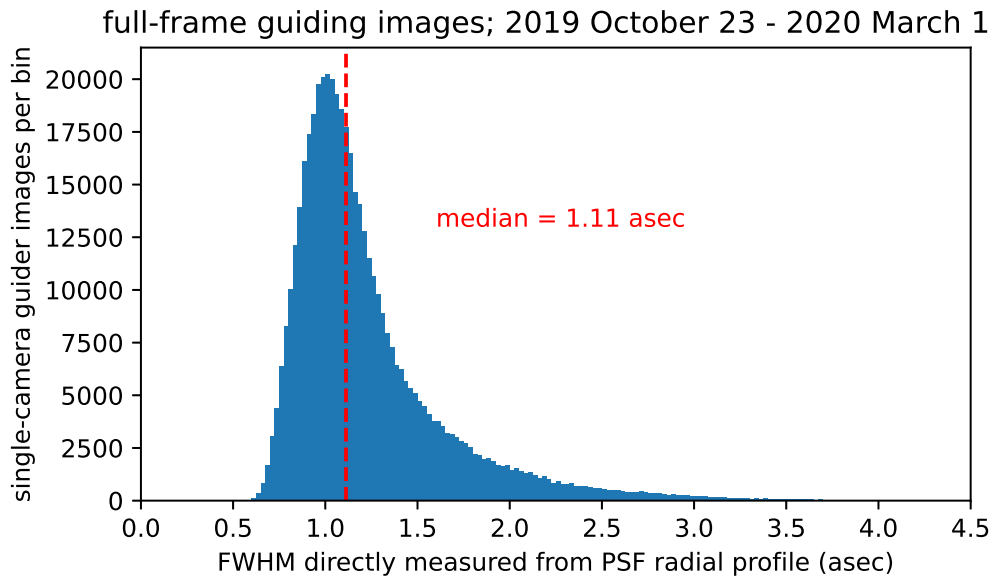}
	\end{tabular}
	\end{center}
   \caption[example] 
   { \label{fig:guiding_fwhm} Histogram of per guider camera FWHM measurements across all of DESI commissioning. Approximately 480,000 single-camera full-frame guider images are included. The FWHM values are directly measured from PSF model radial profiles and therefore do not assume any analytic functional form. The median FWHM value is 1.11$''$, consistent with the pre-DESI Mayall/MOSAIC R-band delivered image quality. We confirmed through visual inspection that a number of images have FWHM values reaching 0.6-0.65$''$, as indicated by the lower tail of this distribution.}
   \end{figure}

For each of the 479,755 retained guider images, we constructed the PSF radial profile and used this profile to directly measure the FWHM (i.e., our FWHM measurements did not assume any functional form). The radial profile construction and corresponding FWHM determination were performed using a port of the relevant subroutines from IDLUTILS ATV\footnote{\url{http://www.sdss3.org/svn/repo/idlutils/tags/v5_5_5/pro/plot/atv.pro}, specifically the atv\_radplotf and atv\_splinefwhm subroutines.}. We estimate a typical uncertainty of 0.07$''$ on our per-camera FWHM measurements, by comparing FWHM measurements of adjacent guider cameras within the same exposure. This estimate is likely best regarded as an upper bound on our per-camera FWHM uncertainty, since adjacent guider cameras have sufficiently large angular separations ($\sim$1$^{\circ}$) that differences in their FWHM values within the same exposure could be due in large part to atmospheric seeing. A histogram of our per-camera FWHM measurements is shown in Figure \ref{fig:guiding_fwhm}. The median FWHM value is 1.11$''$, very similar to the median Mayall/MOSAIC R-band DIQ of 1.17$''$. This agreement is all the more impressive since we are comparing DESI guider images that are $\approx$1.6$^{\circ}$ off axis with the on-axis MOSAIC images. There is a tail of large FWHM values, which is likely due to specific nights when the atmosphere was unusually turbulent and/or the Mayall primary mirror temperature was far different from ambient. We confirmed through visual inspection that a number of images have FWHM values reaching 0.6-0.65$''$, as indicated by the lower tail of the distribution shown in Figure \ref{fig:guiding_fwhm}. Such FWHM $\approx 0.6$-$0.65''$ image quality
is remarkably sharp considering that the DESI guider cameras are 1.57$^{\circ}$ off axis.

% There is also a tail toward small FWHM, reaching as low as $\sim$0.6$-0.65''$. 

%We visually inspected guider images at this end of the distribution, to determine whether they were valid measurements or instead attributable to rare data processing issues. In doing so, we found a number of legitimate cases of guider images with FWHM between 0.6$''$ and 0.65$''$, remarkably sharp considering that the DESI guider cameras are 1.57$^{\circ}$ off axis.

% re: fwhm distribution - first thing i need to 

%\subsection{DESI Guiding Data Set for DIQ Characterization}
%\label{sec:diq_guiding_sample}

%\subsection{FWHM Distribution}

%\subsection{Moffat Parameterization}

% not obvious to me that i need to break this up into any subsections?

\begin{figure} [ht]
   \begin{center}
   \begin{tabular}{c} 
   \includegraphics[height=8cm]{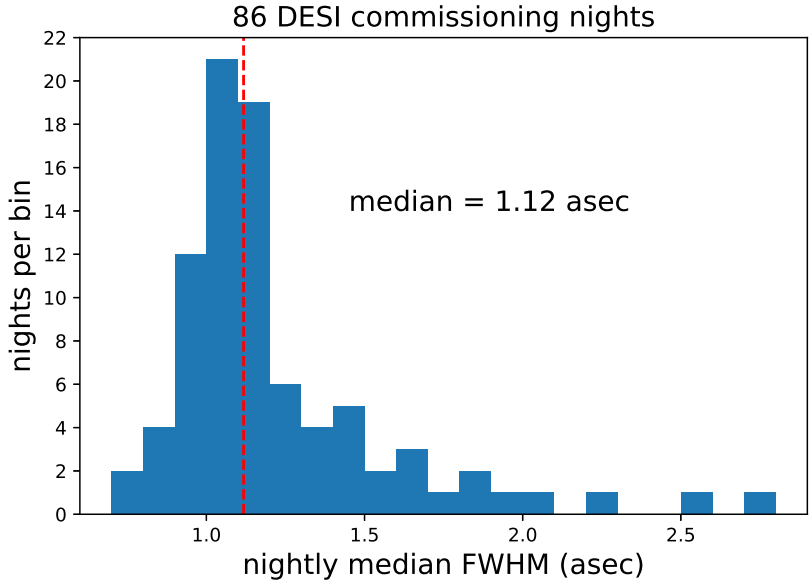}
	\end{tabular}
	\end{center}
   \caption[example] 
   { \label{fig:nightly_fwhm} Histogram of nightly median r-band delivered image quality during DESI commissioning, as measured with the DESI guider cameras. 86 observing nights with at least 600 FWHM measurements per night are included.}
   \end{figure}

% maybe i need an equation in here for the function form of moffat profile ?
Figure \ref{fig:nightly_fwhm} shows a histogram of the nightly median FWHM values across this set of 479,755 individual FWHM measurements. In total 92 observing nights are represented in this sample. Figure \ref{fig:nightly_fwhm} only shows those nights with at least 600 FWHM measurements available. The median nightly FWHM value is 1.12$''$, and the nightly FWHM distribution is generally similar to the per-image FWHM distribution of Figure \ref{fig:guiding_fwhm}. This is consistent with a scenario whereby the seeing varies in large part on a night-to-night basis, primarily due to the degree of atmospheric turbulence, but also to some extent dictated by turbulence associated with the primary mirror's temperature differential versus ambient; this picture was similarly supported by the foregoing Mayall/MOSAIC DIQ study\cite{dey_valdes}. With future observations, we might expect to measure an even better median Mayall/DESI DIQ because  primary mirror cooling was often disabled during DESI commissioning in order to accommodate focal plane positioner testing. During DESI commissioning the primary mirror was a median of 0.44$^{\circ}$ C warmer than ambient with a standard deviation of 1.53$^{\circ}$ C, and the absolute mirror versus ambient temperature differential was $> 1^{\circ}$ C ($> 3^{\circ}$ C) 48\% (6\%) of the time. During DESI survey operations, the primary mirror temperature should better match ambient on average, reducing mirror-induced turbulence.

\begin{figure} [ht]
   \begin{center}
   \begin{tabular}{c} 
   \includegraphics[height=8cm]{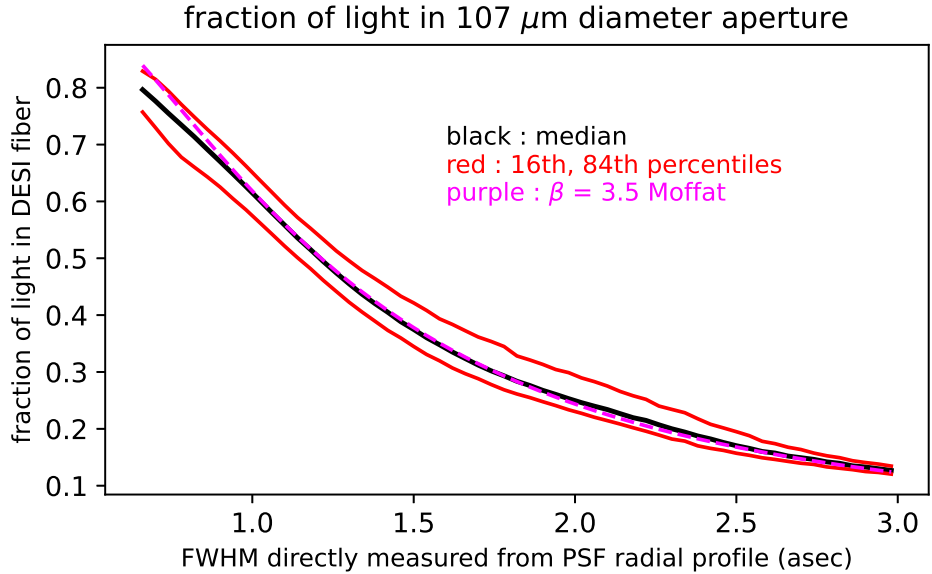}
	\end{tabular}
	\end{center}
   \caption[example] 
   { \label{fig:fracflux} Fraction of flux in an aperture the size/shape of a DESI fiber, as a function of FWHM. The black line provides the median measured trend and the red lines give the corresponding 16th and 84th percentile values. Also overplotted for comparison is a $\beta = 3.5$ Moffat profile (purple dashed line). The measured median trend agrees very well with the $\beta = 3.5$ Moffat profile.}
   \end{figure}

Another way of quantifying the DIQ is to calculate the fraction of each PSF model's flux that would fall within a 107~$\mu$m circular DESI fiber ($f_{fiber}$), assuming that the fiber is perfectly centered on the PSF model centroid\footnote{Our $f_{fiber}$ calculations always use the plate scale at the guider camera locations within the focal plane. The plate scale across DESI's field of view is not constant, and the 107~$\mu$m fiber diameter will correspond to a solid angle that varies as a function of radius from the DESI focal plane center.}. For DESI spectroscopy, this is a more directly relevant DIQ metric than the FWHM, although FWHM is advantageous in terms of ability to compare against prior Mayall/MOSAIC benchmarks. For all 479,755 guider PSF models, we computed the $f_{fiber}$ metric. A plot of the resulting $f_{fiber}$ trend as a function of FWHM is shown in Figure \ref{fig:fracflux}. The moving median (black line) is a near-perfect match to the trend expected for a $\beta = 3.5$ Moffat profile (dashed magenta), except perhaps at very low FWHM values ($\approx$0.8$''$), where the $\beta = 3.5$ curve intersects the $1\sigma$ high measured trendline rather than the moving median. The consistency of the measured trend and Moffat $\beta = 3.5$ behavior indicates that, for the purposes of DESI spectroscopy, the delivered PSF is effectively behaving in the same way as would a Moffat profile with $\beta = 3.5$.

\begin{figure} [ht]
   \begin{center}
   \begin{tabular}{c} 
   \includegraphics[height=4.7cm]{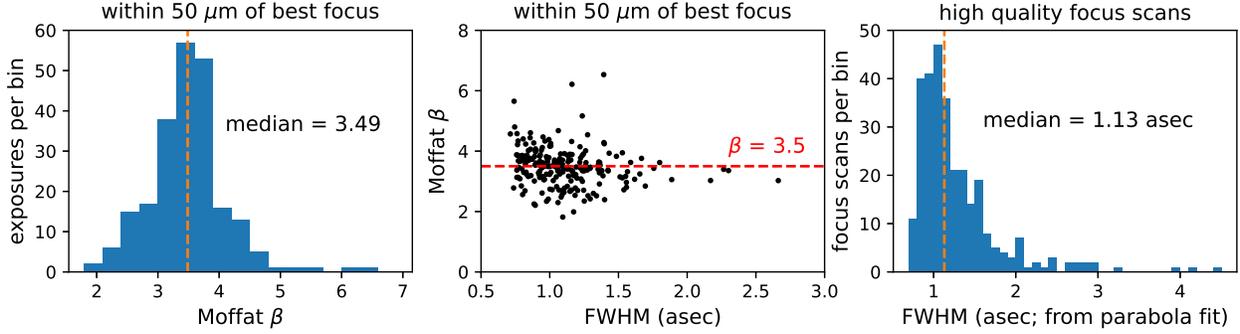}
	\end{tabular}
	\end{center}
   \caption[example] 
   { \label{fig:focus_fwhm_beta} Focus scans offer opportunities to study both the Moffat $\beta$ parameter and the FWHM distribution. Left: distribution of best-fit Moffat $\beta$ values for a sample of guider exposures acquired during focus scans and within 50~$\mu$m of best focus. The median is $\beta$ = 3.49. Center: Moffat $\beta$ as a function of FWHM, for the same sample of exposures in the left panel. There is little dependence of $\beta$ on FWHM. Right: distribution of FWHM values at best focus based on focus sweep parabola fits. The median FWHM value is very similar to that of exposures acquired while guiding (Figure \ref{fig:guiding_fwhm}).}
   \end{figure}

We also implemented the capability to fit each of our two-dimensional PSF models with a Moffat profile of variable $\beta$ parameter, in order to directly measure the Mayall/DESI Moffat $\beta$ parameter distribution. However, we found that the best-fit $\beta$ parameter was subject to large excursions in cases where the guiding data were moderately out-of-focus, corrupting our distribution of $\beta$ values and also the measured trend of $\beta$ versus FWHM. In order to best characterize the Mayall/DESI Moffat $\beta$ parameter, we therefore analyzed a subset of guider camera images that were properly focused. Fortunately, several hundred focus scans (like the one shown in Figure \ref{fig:focus_scan_example}) were accumulated over the course DESI commissioning. The parabolic fits of FWHM versus z for these focus scans provide the best focus z values, such that we can select only those exposures within a small z range about each focus scan's minimum to obtain a clean in-focus data set. The full DESI commissioning focus scan data set contains 227 exposures falling within 50~$\mu$m of their respective focus scan's parabolic fit best z value. For instance, we can see from Figure \ref{fig:focus_scan_example} that exposure number 43855 is one of these 227 cases, since it has z = 389~$\mu$m, which is within only a few $\mu$m of the parabola minimum at z = 392~$\mu$m. For each of these 227 well-focused exposures drawn from focus scans, we again compute pixelized PSF models, radial profile FWHM measurements, and best-fit Moffat $\beta$ values, just as we did for the much larger sample of guiding images. The results are shown in Figure \ref{fig:focus_fwhm_beta}. The left panel is a histogram of the in-focus Moffat $\beta$ parameters. The median value is $\beta = 3.49$, extremely close to the quoted Mayall/MOSAIC value\cite{dey_valdes} of $\beta = 3.5$. The standard deviation of guider camera $\beta$ values is 0.63. The center panel plots Mayall/DESI Moffat $\beta$ as a function of FWHM. No clear trend of $\beta$ with FWHM is discernible, again consistent with the Mayall/MOSAIC findings\cite{dey_valdes}. The rightmost panel shows a histogram of FWHM values determined from the full set of DESI commissioning focus scans, serving as a cross-check on the FWHM distributions presented in Figures \ref{fig:guiding_fwhm} and \ref{fig:nightly_fwhm}. We would expect the distribution of focus scan best FWHM values to be consistent with the distributions of guiding FWHM values and nightly median guiding FHWM values. Indeed, the focus scan FWHM values have a median FWHM of 1.13$''$, very close to the median values from Figures \ref{fig:guiding_fwhm} and \ref{fig:nightly_fwhm}.

% how does my stddev(beta) compare to that of the Dey+Valdes study? or is that info not given in the dey+valdes study?

\begin{figure} [ht]
   \begin{center}
   \begin{tabular}{c} 
   \includegraphics[height=8cm]{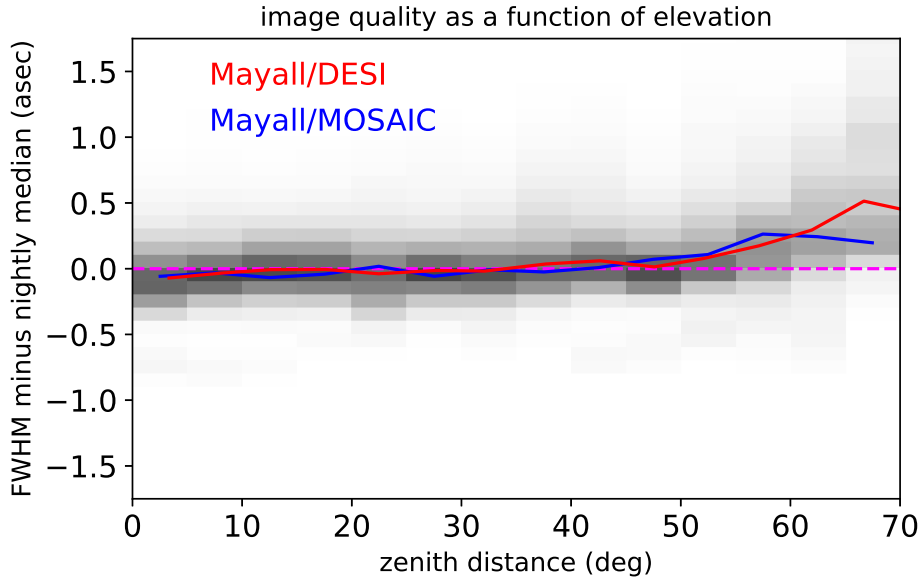}
	\end{tabular}
	\end{center}
   \caption[example]
   { \label{fig:fwhm_vs_zd} Examination of Mayall/DESI delivered image quality as a function of zenith distance. $\sim$480,000 single-camera DESI guider images contribute to this plot, with grayscale showing the density of data points binned by zenith distance and difference between each observation's FWHM and the corresponding nightly median FWHM. Each column of the two-dimensional histogram is normalized so that it sums to unity, since there are far more observations available near zenith than at large zenith distances. The moving median trend of Mayall/DESI image quality versus zenith distance (red line) remains quite flat until reaching $\sim$50$^{\circ}$. The corresponding Mayall/MOSAIC trend (blue line) is very similar. The dashed magenta line denotes image quality independent of zenith distance.}
   \end{figure}

Lastly, we have undertaken a preliminary investigation of the Mayall/DESI image quality as a function of elevation, using the aforementioned sample of 479,755 guiding images. Figure \ref{fig:fwhm_vs_zd} shows the trend of FWHM minus nightly median FWHM as a function of zenith distance. Essentially no trend is seen until zenith distance reaches $\sim$50$^{\circ}$, at which point there becomes a noticeable degradation of the image quality toward higher zenith distances (lower elevations). The Mayall/MOSAIC trend of image quality versus zenith distance\cite{dey_valdes} is also overplotted as a blue line. The Mayall/MOSAIC trend is very similar to the Mayall/DESI trend. We caution that at large zenith distance there is relatively little DESI commissioning data currently available. % , sampling a relatively small number of nights.
 
 % might be good to mention the angular size of each PSF postage stamp cutout?

\section{Conclusion}
\label{sec:conclusion}

Using DESI commissioning data, we have characterized the Mayall telescope's performance in terms of pointing, tracking and delivered image quality. We find that, with its new DESI top end installed, the Mayall telescope is already performing at essentially the same high level as it did in its pre-DESI configuration. In the future, the Mayall/DESI delivered image quality may be further improved by enabling DESI's active focus adjustments, and by collecting more data at relatively low elevation to refine the focus/alignment look-up table in that regime. It will be valuable to repeat our delivered image quality assessment once a sizable data set has been acquired with active focus adjustment, and also once a longer time interval has been sampled (several years versus just five months of DESI commissioning). The favorable Mayall/DESI performance metrics presented in this study bode well for DESI's upcoming five years of cosmology survey operations.

\acknowledgments % equivalent to \section*{ACKNOWLEDGMENTS}       
% couldn't find any IDLUTILS recommended acknowledgment via Google search?
 
 % acks copied from https://www.desi.lbl.gov/acknowledgements/
 % as accessed on 2020 October 17
 
We thank Frank Valdes for valuable feedback that improved this manuscript. This research is supported by the Director, Office of Science, Office of High Energy Physics of the U.S. Department of Energy under Contract No.DE–AC02–05CH1123, and by the National Energy Research Scientific Computing Center, a DOE Office of Science User Facility under the same contract; additional support for DESI is provided by the U.S. National Science Foundation, Division of Astronomical Sciences under Contract No. AST-0950945 to the NSF’s National Optical-Infrared Astronomy Research Laboratory; the Science and Technologies Facilities Council of the United Kingdom; the Gordon and Betty Moore Foundation; the Heising-Simons Foundation; the French Alternative Energies and Atomic Energy Commission (CEA); the National Council of Science and Technology of Mexico; the Ministry of Economy of Spain, and by the DESI Member Institutions. The authors are honored to be permitted to conduct astronomical research on Iolkam Du’ag (Kitt Peak), a mountain with particular significance to the Tohono O’odham Nation.

% References
\bibliography{report} % bibliography data in report.bib
\bibliographystyle{spiebib} % makes bibtex use spiebib.bst

\end{document}